# Enabling Multiple Access for Non-Line-of-Sight Light-to-Camera Communications

Fan Yang, Shining Li, Zhe Yang, *Member, IEEE,* Tao Gu, *Senior Member, IEEE,* and Cheng Qian

**Abstract**—Light-to-Camera Communications (LCC) have emerged as a new wireless communication technology with great potential to benefit a broad range of applications. However, the existing LCC systems either require cameras directly facing to the lights or can only communicate over a single link, resulting in low throughputs and being fragile to ambient illuminant interference. We present HYCACO, a novel LCC system, which enables multiple light emitting diodes (LEDs) with an unaltered camera to communicate via the non-line-of-sight (NLoS) links. Different from other NLoS LCC systems, the proposed scheme is resilient to the complex indoor luminous environment. HYCACO can decode the messages by exploring the mixed reflected optical signals transmitted from multiple LEDs. By further exploiting the rolling shutter mechanism, we present the optimal optical frequencies and camera exposure duration selection strategy to achieve the best performance. We built a hardware prototype to demonstrate the efficiency of the proposed scheme under different application scenarios. The experimental results show that the system throughput reaches 4.5 kbps on iPhone 6s with three transmitters. With the robustness, improved system throughput and ease of use, HYCACO has great potentials to be used in a wide range of applications such as advertising, tagging objects, and device certifications.

✦

## 1 INTRODUCTION

IN recent years, the rolling shutter based visible light communication (VLC) [1] has been a promising technique which uses the Complementary Metal-Oxide-Semiconductor (CMOS) sensor within a digital camera for data reception. It enables the camera to sample optical signals at a much faster rate than the frame rate, and can utilize the pervasively deployed commercial off-the-shelf (COTS) LEDs as transmitters. Since most of the COTS smartphones have CMOS cameras built-in, LCC has shown great potential in short range wireless communications. Furthermore, LCC has some unique features, e.g., it provides a natural way to visually associate the received information with the transmitter's identity, which can be used in indoor localization, augmented reality, etc.

LCC can be classified into two categories. One is line-of-sight (LoS), i.e., the camera directly faces to the LED [2]. The other one is non-line-of-sight (NLoS), i.e., the camera observes the reflected optical signals [3], as illustrated in Fig. 1. However, each of the kinds has hit its bottleneck of throughput. The throughput depends on several factors including the signal frequency and the camera features, mostly importantly on the region of interest (RoI) [4]. The NLoS LCC typically has higher throughput because the optical signals occupy the whole image, while the optical signals only occupy part of the image via the LoS link unless the camera is very close to the LED. However, the received light strength of an NLoS link is attenuated significantly due to the diffuse reflection, which increases the demodulation error. Moreover, optical signals from other illuminants are mixed by the reflector, which causes significant interferences. Therefore, it is easy to enable multiple access for LoS LCC with a cost of lower throughputs, while NLoS LCC has higher throughputs but hard to enable multiple access.

Enabling multiple access for NLoS LCC is challenging for four main reasons: (i) The information transmitted from each LED is hard to be extracted because the entire image is filled with the mixed reflected optical signals. (ii) The captured optical signals are mixed with the environment background and image noise introduced by camera hardware. In LCC, the images are usually captured with high ISOs and short exposure durations, which causes substantial random noise, including chroma noise and luminance noise. (iii) Paramount to the practical implementation of LCC is ensuring high-quality lighting that is satisfactory to human, which limits the design space of the optical waveform. (iv) LCC is a one-way communication link, via which the transmitter cannot get any feedback from the receiver. It brings intrinsic difficulties to implement the unsynchronized communication in realtime.

In this paper, we propose a HYbrid light-to-CAmera COmmunication (HYCACO) system. Different from many existing approaches, HYCACO works in more realistic indoor luminous environments (i.e., multiple light sources and natural illumination), where multiple LEDs are coordinated to emit square wave signals simultaneously. We embed the information in the phase of the square waves, which is called phase-shift keying (PSK). Different from the conventional PSK, we propose a new scheme, hybrid PSK (HPSK), where the waveform emitted from each LED adopts different orders of PSKs. Therefore, the received signals can be considered as a combination of square waves with different properties, e.g., phase and/or frequency. In particular, when a CMOS camera obtains an image with

---

- F. Yang, C. Qian, S. Li and Z. Yang are with the School of Computer Science and Engineering, Northwestern Polytechnical University, Xi'an China, 710129. E-mail: {craftsman, infinite}@mail.nwpu.edu.cn, {lishining, zyang}@nwpu.edu.cn.
- T. Gu is with School of Computer Science and IT, RMIT University, Australia. Email: tao.gu@rmit.edu.au

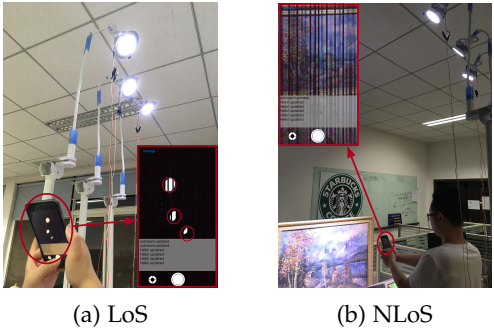

(a) LoS      (b) NLoS

Fig. 1: Comparison between LoS and NLoS light-to-camera communications.

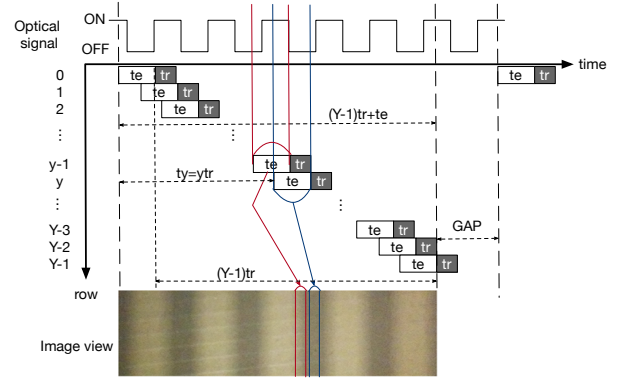

Fig. 2: The optical signal is a square wave. $t_e$ is the exposure duration of the captured image and $t_r$ is the readout duration of the camera.

respect to the reflector illuminated by the LEDs, it contains bright and dark bands corresponding to the mixed optical signals. In order to extract information from the mixed signals, we propose a SUperimposed Rect-wave Division (SURD) algorithm.

We further exploit the rolling shutter effect and figure out the relationship between the camera settings and the features of optical signals. We propose a solution to recover the optical signal by capturing two images with different exposure durations but the same exposure value (EV). We use a simple preamble to help to sample hundreds of judging points from millions of pixels in the image. Thus, HYCACO has high computational efficiency and can provide realtime responses with the COTS smartphones. Besides, We address the symbol loss problem, which is caused by the unsynchronized communication channel, with the Luby transform codes (LT codes) [5].

We have implemented a prototype system to evaluate the performance of HYCACO. We employ an Arduino UNO board to act as the modulator to control up to 7 LEDs as the transmitter. At the receiver end, we develop an iOS application and an Android application to verify the performance on iPhone 6s and Nexus 5, respectively. We conduct the extensive experiments under various camera settings and environments to evaluate the system performance comprehensively. The experimental results have demonstrated the efficacy of the proposed scheme.

In summary, this paper makes the following contributions:

- To our knowledge, HYCACO is the first NLoS LCC system which enables multiple access. We have implemented a prototype system and test the performance of HYCACO in different scenarios. The extensive experimental results demonstrate that HYCACO can achieve a throughput of $4.5kbps$. This is a significant improvement compared with other state-of-the-art LCC systems.
- We propose a new modulation scheme, HPSK, to transmit messages from multiple LEDs and also propose the corresponding algorithm, SURD, to demodulate the signals.
- We first present the relationship between the exposure settings and the features of optical signals, and come up with the optimal optical frequency and camera exposure duration selection strategy.
- We propose a simple solution to separate the captured optical signals from the complex image background, which improves the robustness of the LCC systems.

## 2 PRELIMINARY

Before presenting the proposed HYCACO scheme, we present the preliminary about the rolling shutter mechanism and the image capture model of the mainstream CMOS cameras. Then we describe the characteristics of unsynchronized communication in LCC.

### 2.1 Image Capture Model

The electronic rolling shutters have been widely employed by most of the smartphones which have CMOS cameras built-in. When a CMOS camera captures photos or videos, it does not expose every pixel of the entire image all at once. Instead, each sensor array of pixels in the image is triggered by row (shown in Fig. 2). The exposure duration of a pixel row is shifted by a fixed amount of readout duration. It means that a pulsing light will illuminate only some rows of the pixels at a time, resulting in the alternately dark and bright bands in the image. We can detect the frequency of these bands in the image, and thus infer the frequency of the pulsing light. Therefore, it can act as a sampling process to the optical signal with much higher sampling rate than the frame rate of the camera. As a result, this effect allows us to record the flicker pattern by taking spatio-temporal images with an unaltered digital camera, where different patterns can be used to represent different symbols.

Conceptually, the image captured by the camera can be thought of as having two layers: the texture layer and the signal layer. HYCACO consists of several ON-OFF keying (OOK) modulated LEDs as the transmitters and a CMOS sensor as the receiver. The luminance emitted from an LED is $L$, and we use the input signal $s_i(t)$ to control the state of the LED, where $i$ denotes the serial number of the LED. If $s_i(t) = 1$, the LED is ON; if $s_i(t) = 0$, the LED is OFF. Thus, the illuminance of light falling on the camera sensor is

$$e(t) = E + \sum H_i(0)s_i(t)L, \quad (1)$$

where $H_i(0)$ is the channel DC gain [6] and $E$ is the non-flickering lights (such as sunlight). Let $r(x, y, t)$ be the radiance incident at sensor pixel $(x, y)$ at time $t$. The radiance $r(x, y, t)$ can be factorized into spatial and temporal components:

$$r(x, y, t) = l(x, y)e(t), \qquad (2)$$

where $l(x, y)$ is the amplitude of the temporal radiance profile at pixel $(x, y)$ and is determined by the image background. The measured brightness value of a pixel at $(x, y)$ in the image is [7]

$$\begin{aligned} i(x,y) &= k\, l(x,y) \int_{-\infty}^{\infty} f(t_y - t) e(t) dt + n(x,y), \\ &= \underbrace{kl(x,y)}_{\text{texture layer}} \times \underbrace{(f * e)(t_y)}_{\text{signal layer}} + n(x,y), \end{aligned} \qquad (3)$$

where $k$ is the sensor gain, $t_y$ is the temporal shift for a pixel in row $y$, and $t_y = yt_r$ (as illustrated in Fig. 2). $t_r$ is the readout duration. $f(t)$ is the shutter function. If the pixels in row $y$ capture light at time $t$, $f(t) = 1$; otherwise, $f(t) = 0$. $n(x, y)$ is the image noise. Since the signal layer is unidimensional, we perform analysis on vertical sum images—that is, $i(y) = \sum_x i(x, y)$, $l(y) = \sum_x l(x, y)$ and $n(y) = \sum_x n(x, y)$. Then, Equation (3) can be written as

$$i(y) = k\, l(y) \times (f * e)(t_y) + n(y). \qquad (4)$$

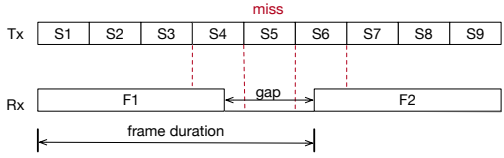

Fig. 3: Mixed symbol and symbol loss due to unsynchronization.

### 2.2 Unsynchronized LCC Channel

The characteristic of a camera's discontinuous receiving and the diversity of cameras lead to an unsynchronized LCC channel. Such an unsynchronized communication channel is very likely to experience the mixed symbol frame and the symbol loss problems. RollingLight [2] demonstrates these problems under different unsynchronized scenarios. The optical signals emitted from the LEDs are continuous, but the camera receives the signals frame by frame. A camera does not expose at all time in a frame duration. There exists a time gap between the end time of the exposure of the last row and the start time of the next frame, as illustrated in Fig. 2.

For example, as shown in Fig. 3, there exists a time gap between the end of exposure in frame $F1$ and the start of exposure in frame $F2$. Frame $F1$ receives a mixture of symbol $S1$, $S2$, $S3$, and part of $S4$. When the length of the gap is longer than the symbol duration, some symbols may be completely lost, e.g., symbol $S5$. In this case, symbol $S4$, $S5$, $S6$ will not be extracted by the receiver. Different frame rates cause different levels of unsynchronizations, which lead to different symbol loss ratios [8].

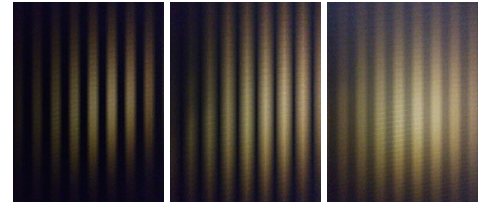

(a) Captured images

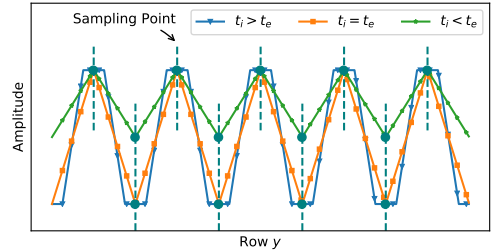

(b) Simulated signal layers

Fig. 4: The brightnesses of the bands change gradually due to the rolling shutter effect. $t_i$ is the minimum pulse duration, and $t_e$ is the exposure duration.

## 3 CARRIER FREQUENCY SELECTION

In this section, we study the lower and upper frequency limits of the transmissions, and how exposure duration impacts the performance of the LCC systems. So we can choose the optimal optical frequencies and exposure duration for HYCACO.

### 3.1 Lighting Requirements

Without loss of generality, we assume that LCC is used for both lighting and communication. Hence, the LEDs used in the LCC system should meet the lighting requirements for human. When the carrier frequency is smaller than eye's temporal resolution, called critical flicker frequency (CFF), flicker happens [9]. Typically, human eyes are able to resolve up to 50Hz to luminance flicker and 25Hz to chromatic flicker [10]. Although human eye has a cutoff frequency in the vicinity of 50Hz, some studies have shown that long-term exposure to higher frequency (unintentional) flickering (in the 70 to 160 Hz range) can also cause malaise, headaches, and visual impairment [11]. Besides, the perceived brightness of an LED varies proportionally to the average duty cycle of its flicker pattern. Therefore, the duty cycle in a CFF cycle duration should be unchanged too. We set the lower frequency limit of the optical signals to 200 Hz and modulate the waveforms with a duty cycle of 50%.

### 3.2 Pulse Duration vs. Exposure Duration

The duration of the optical signals recorded in one frame is proportional to the width of the RoI and the readout duration $t_r$. One advantage of NLoS compared to LoS is it can easily amplify the RoI to the full width of the receiver. As illustrated in Fig. 2, the camera captures an image with an exposure duration $t_e$. Let $Y$ denote the width (number of pixel rows) of the image. The duration for the imager to open and allow photons to enter is $(Y-1)t_r + t_e$. However, the signal layer of the image is the convolution of the shutter function and the optical signal. Hence the duration of the signal recorded in the image is $(Y-1)t_r$.

Although the exposure duration makes no reference to the signal duration of the image, it affects the gradient pattern of the bands in the image. For most cameras of smartphones, the shutter function can be considered as a window function, and the window length is the exposure duration. The square waves are convoluted with the shutter function which causes the gradient effects of the bands, though this may not be obvious to the naked eyes. The capturing can be classified into three circumstances according to the relationship of the minimum pulse duration $t_i$ and the exposure duration $t_e$, i.e., $t_i > t_e$, $t_i = t_e$, and $t_i < t_e$, as shown in Fig. 4a. We can see that under the same pulsing LED, the length of gradient increases as the exposure duration increases in the captured images. The simulated signal layers of the images in Fig. 4a are illustrated in Fig. 4b. We can see that the shutter function deforms the square wave by convolution, but it does not change the frequency of the original waveform, and when $t_i \geq t_e$, the amplitude of the signal layer is proportional to the original waveform, too. Only when $t_i < t_e$, the signal layer losses the spatial details of the original waveform. Therefore, as long as $t_i \geq t_e$, there exists a set of sample points in the signal layer that can fully describe the original waveform, as illustrated in Fig. 4b.

Furthermore, there are some physical constraints of cameras placed on the reception of the optical signal. Cameras on the market usually have different $t_r$. We propose a simple method to calibrate $t_r$ by sending a known preamble. We measure several phones' $t_r$ and the supported range of exposure duration. We find $t_r$ is smaller than the minimum exposure duration that the OS allows the user to set. $t_r$ is usually from several microseconds to a dozen microseconds. Theoretically, the upper frequency limit is $1/2t_r$ Hz. However, when $t_i < t_e$, we cannot infer the accurate spatial detail of the original waveform, and when $t_i \ll t_e$, the signal layer is approximately constant. In our proposal, HYCACO needs both the spatial and temporal details of the waveforms to perform the demodulation. Thus, we set $t_i = t_e$ to obtain the maximum achievable throughput.

## 4 HYCACO DESIGN

The architecture of HYCACO is shown in Fig. 5. Several COTS LEDs each connected to a transistor switch circuit are employed as the transmitters. A microcontroller encodes the input data to *ON-OFF* symbols and dispatches the symbols to the circuits. Thus, the data waveforms are modulated onto the instantaneous power of the optical carriers. A smartphone with a built-in CMOS camera is employed as the receiver. First, the smartphone continuously takes images of the reflector (a rough surface) and calculates the signal layer of each image. Second, the signal layers are demodulated to several sequences of N-ary symbols according to the transmitter number. Third, we decode the symbol sequences to binary data packets. Finally, we combine the data packets to retrieve the full input message.

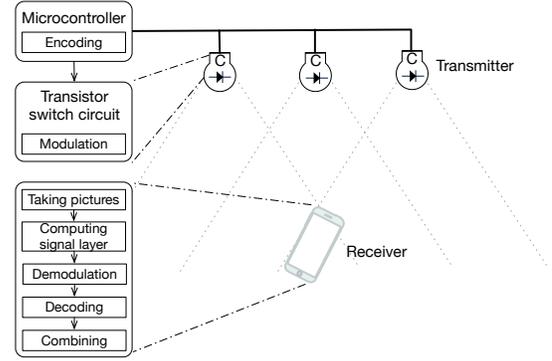

Fig. 5: Shows the architecture of HYCACO. The transmitters are several temporally modulated LEDs, and the receiver is a rolling shutter camera.

### 4.1 Modulation and Encoding Scheme with Multiple Access

The inspiration of our multiple access scheme comes from Orthogonal Frequency Division Multiplexing (OFDM). The waveform emitted from each LED can be considered as a subcarrier. Thus, the optical carrier is

$$s(t) = \sum s_i(t), \quad (5)$$

where $i$ denotes the serial number of the subcarrier/LED. In our prototypes, the transmitters share the same microcontroller and are thus inherently synchronized. By decoding the messages modulated on each subcarrier, the receiver can communicate with each transmitter, respectively. For the convenience of evaluation, we let each transmitter sends a piece of input data, and the receiver combines the pieces to retrieve the full input data.

Our multiplexing technique is much simpler than OFDM. Here are three differences: (i) The frequency is a sine wave in OFDM, while it is a square wave in HYCACO. (ii) The subcarrier itself is not useful in transmitting the information in OFDM, while it conveys information by changing its phase in HYCACO. (iii) Unlike OFDM, HYCACO has no intersymbol interference (ISI) problem, because light suffers less from multipath effect than RF and the frequency of modulation is below $1/2t_r$ Hz (usually less than 100 kHz).

In this paper, a Hybrid PSK (HPSK) scheme is proposed for multi-carrier modulation. As the amplitude of each subcarrier is invariant, to increase the number of distinct symbol changes, we let each frequency adopts its allowed highest order in PSK. The allowed highest order is decided by the period and the time granularity. In the end of Section 3.2, we let the minimum pulse duration $t_i$ equals to the exposure duration $t_e$ to obtain the maximum achievable throughput. Thus, $t_e$ is the time unit, denoted by 1. Let $T$ denote the symbol duration in the unit of $t_e$, and let the subcarrier serial number $i$ represent the number of cycles in one symbol. The period of subcarrier $i$ is $T/i$, and it adopts $T/i$ order of PSK. Fig. 6 illustrates an example of HPSK with two LEDs. Thus, subcarrier $i$ can be expressed as follows:

$$s_i(t) = \begin{cases} 1, & 0 \leq t \bmod \frac{T}{i} + \frac{S_i}{T} < \frac{T}{2i} \\ 0, & \frac{T}{2i} \leq t \bmod \frac{T}{i} + \frac{S_i}{T} < \frac{T}{i} \end{cases}, \quad (6)$$

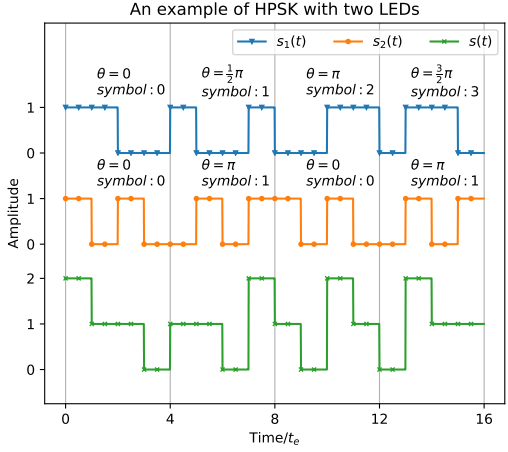

Fig. 6: Shows an example of HPSK modulation with two LEDs. $\theta$ is the phase of the waveform. *symbol* is a N-ary number where N is the order of PSK. The numbers on the Y-axis represent the states of the LEDs, e.g., 2 means two LEDs are ON, 1 means one LED is ON, and 0 means all LEDs are OFF.

where $S_i$ denotes the symbol it represents.

First of all, we need to choose the minimum symbol duration $T$ according to the number of transmitters. Subcarrier $i$ has the following properties:

- The optional range of the frequency is from 200 to $1/2t_e$ Hz.
- $i \in \mathbb{N}$, where $\mathbb{N}$ denotes natural numbers.
- $i \leq T/2$.
- $T/i \in \mathbb{N}$.

Let $\mathbb{I}$ denote the set of subcarrier subscripts and $\mathbb{D}$ denote the set of divisors of $T$. we can derive that $\mathbb{I} \subseteq \mathbb{D} - \{T\}$. Let $I$ denote the transmitter number, and $I = |\mathbb{I}|$. The proper $T$ is

$$\min T \quad \text{s.t.} \ |\mathbb{D} - \{T\}| \geq I. \tag{7}$$

Secondly, we encode the input data to N-ary symbols. The encoding scheme is as follows:

1) Convert the input data to binary.
2) Divide the binary sequence into $I$ sequences, each of which with a length of a multiple of $log_2(T/i)$.
3) Convert the divided sequences to N-ary sequences correspondingly, where $N = T/i$.

Lastly, we distribute these N-ary sequences to the corresponding transmitters.

Let us use an example to demonstrate the encoding and modulation process. As illustrated in Fig. 7a, we use two transmitters to send the message *Hello!*. The modulated optical signal is shown in Fig. 7b.

### 4.2 Signal Recovery

An NLoS link typically has a low signal-to-noise ratio (SNR) because of the complex image background and the diffuse reflection. We tackle these challenges by taking a long exposure image which has the same EV as that of the short exposure images. EV is a number that represents a combination of a camera's exposure duration, ISO, and f-number. The relationship is given by the exposure equation prescribed by ISO 2720:1974[1]. As the f-number is fixed in smartphones, the short is captured with a high ISO and the long is captured with a low ISO. The two images are captured with the same image background and are motion blur-free. The two images are given as

$$i_{short}(y) = k_{short}l(y) \times (f_{short} * e)(t_y) + n_{short}(y), \tag{8}$$

$$i_{long}(y) = k_{long}l(y) \times (f_{long} * e')(t_y) + n_{long}(y), \tag{9}$$

where $k$ is the sensor gain which can be adjusted by the ISO, $l(y)$ is the image background, $f(t)$ is the shutter function, $e(t)$ is the illuminance of light falling on the sensor, which is the sum of received optical signals.

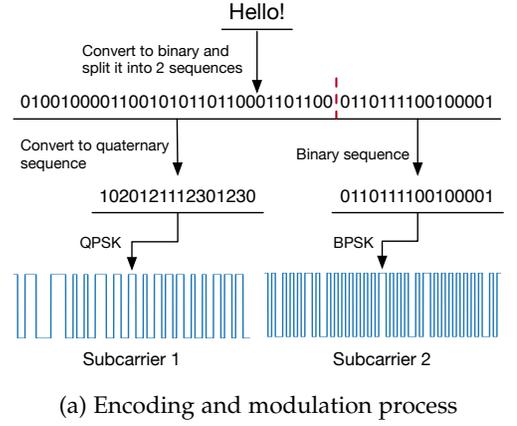

(a) Encoding and modulation process

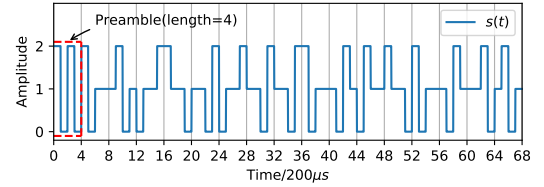

(b) Input Signal

Fig. 7: An HYCACO encoding and modulation example which uses two transmitters to send the message *Hello!*.

The long exposure can be approximated as the texture layer of the short exposure. $f_{long}$ is chosen so that it is significantly longer than the period of the temporal signal, thus $(f_{long} * e')(t_y) \approx K$, where $K$ is a constant. Images captured with the same EV will present the same scene luminance [12], thus $k_{short} = K k_{long}$. Because the long one is captured with a low ISO, its image noise is much smaller than the short one's. After summing the intensities along each image row, $n_{long}(y)$ could be negligible. Let $g(y)$ denote the signal layer of the short exposure. Thus,

$$g(y) \approx \frac{i_{short}(y)}{i_{long}(y)}. \tag{10}$$

As illustrated in Fig. 8, the texture layer may not look the same in the two images. It is because that the short one is captured with a high ISO and thus introduces more image noise. The long exposure only needs to be captured once at the beginning of the reception. As long as the communication time is long enough, the extra cost of capturing the long exposure could be negligible.

---

1. $\frac{N^2}{t} = \frac{LS}{K}$, where $N$ is the relative aperture (f-number), $t$ is the exposure time ("shutter speed") in seconds, $L$ is the average scene luminance, $S$ is the ISO arithmetic speed, $K$ is the reflected-light meter calibration constant.

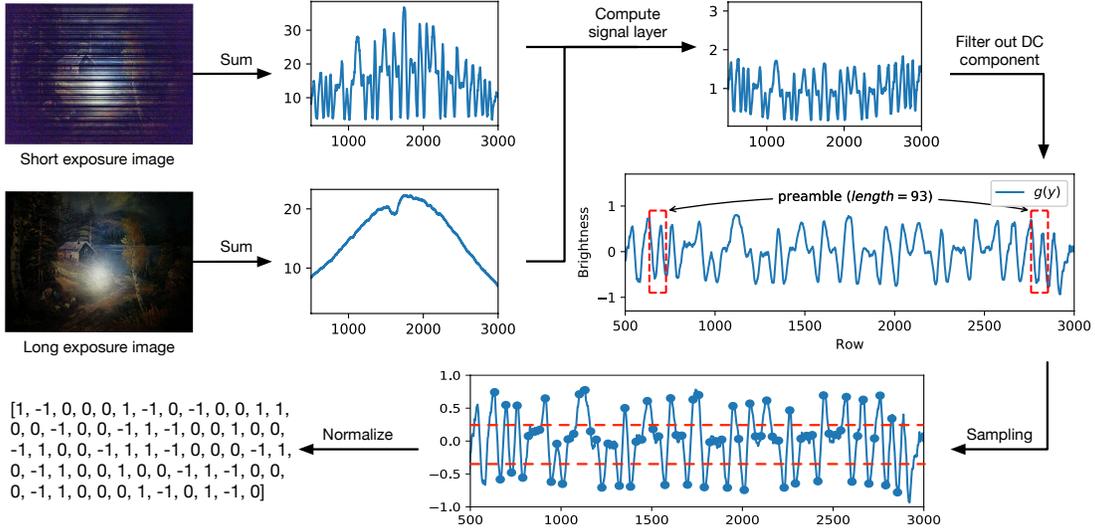

Fig. 8: Shows the process of signal recovery. The long exposure image has the same EV as the short one's. The result is a sequence which represents the illumination levels.

The LEDs are located at different locations and thus have different distances and irradiance/incidence angles to the reflector. The channel gain of each LED-reflector link is different. The channel attenuations can be compensated by detecting the frequencies of the subcarriers. The compensation algorithm will be addressed in our future work. Here, we just assume the channel gains of the links are approximately equal. Thus, equation (1) can be written as

$$e(t) = ILH(0)s(t) + E, \quad (11)$$

where $ILH(0)$ is a constant. Non-flickering component $E$ can be filter out by a DC filter, as illustrated in Fig. 8.

We extract the packets by detecting the preambles in the signal layer. The transmitter sends a known preamble at the beginning of a packet, as shown in Fig. 7b. In the signal layer, the preamble is convolved with the shutter function to form one and a half cycles of a triangle wave, which starts at a peak. The extracted signal layer of the example in Section 4.1 is illustrated in Fig. 8. The preamble also gives us the information of the sampling period. Let $n_p$ denote the width of the preamble in the signal layer. The width proportional to one unit time in the signal layer is $n_p/3$, denoted by $n_i$. $n_i$ can be used to estimate the readout duration $t_r$, i.e., $t_r = t_i/n_i$. We use $n_i$ as the sampling period. Then, we normalize the sampling result to get an illumination level sequence, denoted by $g[k]$, as illustrated in Fig. 8. According to Nyquist–Shannon sampling theorem, $g[k]$ can fully describe $s(t)$.

### 4.3 Demodulation and Decoding

We propose a superimposed rect-wave division algorithm to divide the optical signal into a set of square waves. The Fourier series of Equation (6) is

$$s_i(t) = \frac{1}{2} + \frac{2}{\pi} \sum_{n=1}^{\infty} \frac{sin\left(i(2n-1)t + \theta\right)}{2n-1}, \quad (12)$$

where $t \in [0, T)$. The first sinusoid component ($n = 1$) is the fundamental frequency which has the same frequency and phase as the square wave. When the continuous signal $s_i(t)$ is sampled at the inverse of one unit time Hz, we get the discrete form of $s_i(t)$,

$$x_i[k] = \frac{1}{2} + \frac{2}{\pi} \sum_{n=1}^{T/2i} \frac{sin\left(i(2n-1)2\pi(\frac{k+S_i}{T})\right)}{2n-1}, \quad (13)$$

where $k = 0, 1, 2, \cdots, T-1$. Thus, the discrete form of the optical carrier is $x[k] = \sum x_i[k]$. By taking a real discrete Fourier transform (DFT) of both sides, we get the frequency domain,

$$X[\omega] = \sum X_i[\omega], \qquad i \in \mathbb{I}, \quad (14)$$

where $\omega = 0, 1, 2, \cdots, T/2$. $\angle X_i[i]$ is the phase of the $i^{th}$ subcarrier. The frequency bins $X_i[\omega]$ represent the harmonics which construct subcarrier $x_i$, where $\omega$ is the cycle number of the harmonic. We can see that $X_i[\omega] = 0$ when $\omega \neq i(2n-1)$ and $\omega \neq 0$, hence $X_1[1] = X[1]$ and $X_i[i] = X[i] - \sum_{j=1}^{j=i-1, j \in \mathbb{I}} X_j[i]$. The demodulation and decoding process is expressed in Algorithm 1.

### 4.4 Dealing with Unsynchronized Communications

We address the symbol loss problem with LT codes. LT codes employ a particularly simple algorithm based on the XOR to encode and decode the message. We encode the input data to LT codes before the HYCACO encoding process. The process of generating an encoding packet is easy to describe:

1) Divide the input data into $n$ blocks of roughly equal length.
2) Randomly choose $d$ blocks, where $1 \leq d \leq n$ and the degree $d$ is a pseudorandom number.
3) The value of the encoding symbols is the XOR of the $d$ blocks, i.e., $M_{i_1} \oplus M_{i_2} \oplus \cdots \oplus M_{i_d}$, where $M_i$ is the $i^{th}$ packet and $\{i_1, i_2, \ldots, i_d\}$ are randomly chosen indices of the $d$ blocks.

**Algorithm 1** SUperimposed Rect-wave Division (SURD)

**Input**: $g[k]$, $\mathbb{I}$
**Output**: $S[i]$
$\mathbb{X} \leftarrow$ short-time Fourier transform (STFT) of $g[k]$
**for** each $X \in \mathbb{X}$ **do**
    **for** each $i \in \mathbb{I}$ **do**
        **if** $i = 1$ **then**
            $X_i[i] \leftarrow X[i]$
        **else**
            $X_i[i] \leftarrow X[i] - \sum_{j=1}^{j=i-1, j \in \mathbb{I}} X_j[i]$
        **end if**
        $S[i] \leftarrow \frac{\angle X_i[i] \times T}{2i\pi}$
        Calculate $x_i$ via Equation (13)
        $X_i \leftarrow$ Real DFT of $x_i$
    **end for**
**end for**

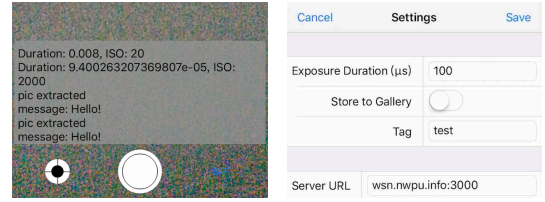

(a) Capturing      (b) Settings

Fig. 10: GUI of receiver app

TABLE 1: Parameters of The Receiver

|  | Exposure Duration ($\mu s$) | ISO | Image Resolution ($X \times Y$) | Frame Rate (fps) | Readout Duration ($\mu s$) |
|---|---|---|---|---|---|
| iPhone 6s | 13-333333 | 23-1840 | 640 × 480-4032×3024 | 3-240 | 6.45 |
| Nexus 5 | 13-866975 | 100-10000 | 640 × 480-3264×2448 | 7-30 | 12.5 |

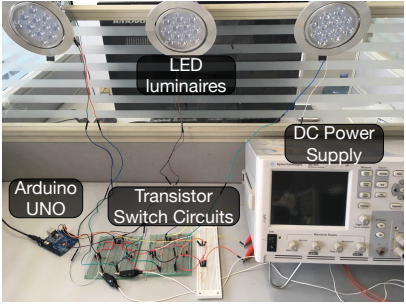

Fig. 9: Experimental equipments of the transmitters

4) A prefix is appended to the symbols defining the list of indices and the total blocks $n$ in the input data.

We perform the HYCACO encoding and modulation process on these encoded packets. The receiver keeps extracting packets from the captured images. If a packet is of degree $d > 1$, it is first XORed against all the decoded packets in a message queuing area, then stored in a buffer area if its reduced degree is greater than 1. When a new packet of degree $d = 1$ is received or reduced, it is moved to the message queuing area and matched against all the packets in the buffer. When all $n$ packets have been moved to the queuing area, the received data has been successfully decoded.

## 5 EVALUATION

In this section, we first evaluate the achievable throughput of HYCACO with different prototype settings, such as smartphone model, transmitter number, packet duration and frame rate. Then we choose the proper settings according to the achievable throughput evaluation results to evaluate the realistic performance under different capturing geometry, illuminance and exposure duration.

### 5.1 Experiment Setup

The transmitters of our hardware prototype consist of a DC power supply, an Arduino UNO, and several COTS LEDs each connected with a transistor switch circuit board (shown in Fig. 9). The transistor switch circuit boards are used to amplify the signal to a proper voltage level for the LEDs. The circuit boards are connected to the Arduino UNO, the microcontroller, which accepts the input data and generates *ON-OFF* symbols. The max forward voltage and current for the LED is 40 V and 350 mA, respectively. These LEDs illuminate a rough surface painting which is the reflector. The reflector is 2.1 m away from the nearest LED. The measured illuminance in the room (all LEDs are off) is about 50 lux.

On the receiver side, we test two devices, iPhone 6s and Nexus 5, and build two apps for iOS and Android. As illustrated in Fig. 10b, users can set the exposure duration $t_e$ equal to the time unit $t_i$. After pressing *Save* button, the app automatically set the ISO of the short exposure to the maximum, and compute the corresponding exposure duration and ISO of the long exposure. We make the exposure duration of the long one hundred times of the short's. Therefore, the long exposure is captured with a small ISO and a long exposure duration, which makes the long exposure nearly flicker-free and noise-free. The setting ranges and the calibrated readout durations of the receivers are illustrated in Table 1. This app has two working modes: (a) process and demodulate the image within the phone; (b) process the image within the phone and upload the sampling result $g[k]$ to a server, then the server demodulates $g[k]$ and sends the retrieved message back to the phone, as illustrated in Fig. 10a. Mode (b) is designed for single frame communication such as indoor positioning.

### 5.2 Achievable Throughput

We evaluate the achievable throughput by choosing the frames without reception error. Since the frame rates of different cameras are slightly different, we address this problem and come up with a definition of "frame throughput" quantified using the bits per frame unit (symbol: "bit/f"). We set the time unit to 100 $\mu s$ and set the resolution to the maximum.

#### 5.2.1 Throughput vs. Transmitter Number
According to Equation (7), the more transmitters, the longer duration the transmitter needs to send a symbol. Therefore,

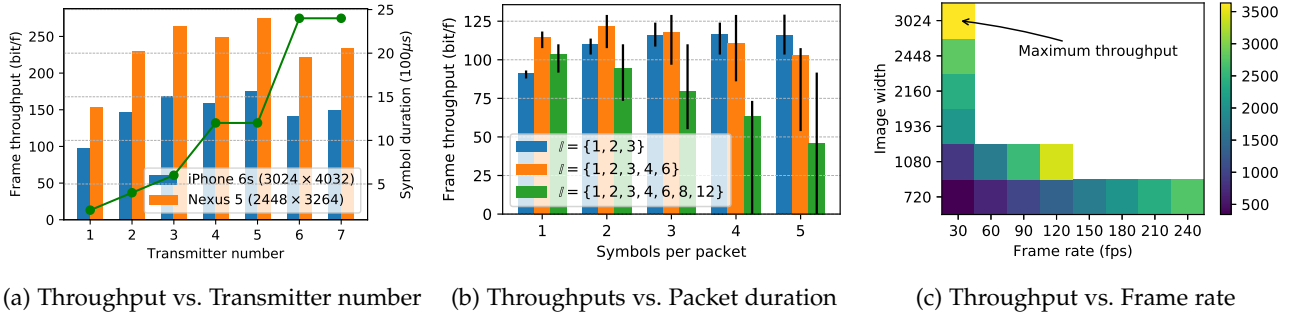

(a) Throughput vs. Transmitter number  (b) Throughputs vs. Packet duration  (c) Throughput vs. Frame rate

Fig. 11: The achievable throughput. *Frame throughput* is the amount of data decoded per frame.

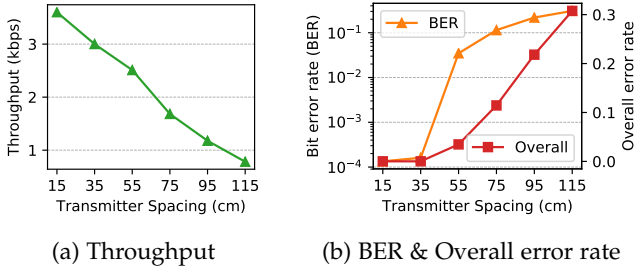

(a) Throughput  (b) BER & Overall error rate

Fig. 12: Throughput, BER and overall error rate with a varying transmitter spacing.

the number of transmitters determines the symbol duration, the symbol duration determines the granularity of a packet, and different levels of granularity cause different symbol loss ratios. If no symbol is lost (no signal in the image is discarded), given the image resolution and readout duration, we can simulate the frame throughputs with different numbers of transmitters for transmission. The results are illustrated in Fig. 11a. We can see that HYCACO reaches the highest frame throughput when the number of transmitters is 5, symbol duration $T$ is 12 units of time, and the subcarrier numbers $\mathbb{I}$ are $\{1, 2, 3, 4, 6\}$.

The frame throughput on Nexus 5 is higher than on iPhone 6s, although images captured by iPhone 6s have higher resolution. This is because the readout duration of iPhone 6s is much smaller than that of Nexus 5. According to our experiments, the readout duration of iPhone 6s and Nexus 5 is 6.45 $\mu s$ and 12.5 $\mu s$, respectively. Although on Nexus 5 can theoretically achieve higher throughput, images captured by Nexus 5 contain more noise, and iPhone 6s has a broader range of frame rate. Therefore, we use iPhone 6s as the receiver for the rest of the evaluation.

#### 5.2.2 Throughput vs. Packet duration

Even if we assume that there is no reception error, the throughput is still variable due to unsynchronization. We need a preamble preceding each packet to extract the packets from the signal layer. The preamble takes an extra cost of 4 units of time for the transmitter to send a packet. Packet duration equals the sum of the preamble duration and the product of the symbol duration and symbols per packet. Signals before the first detected preamble and after the last detected preamble are discarded, which causes symbol losses. Fig. 11b shows how the packet duration affects the frame throughput. We can see the variance increases as the packet duration increases. When the packet duration is small, the preambles take too many shares in the signal layer, which reduces the throughput; when the packet duration is big, the discarded signals might be too long which increases the variance.

In real deploy environments, the LEDs are installed at different locations, which leads to channel gain differences of the LED-reflector links. Without channel compensation, more transmitters would cause more reception error. We choose four symbols per packet and the subcarrier numbers $\{1, 2, 3\}$ for the rest of the evaluation, because this combination achieves nearly highest average throughput with fewer transmitters.

#### 5.2.3 Throughput vs. Frame Rate

We check the impact of the camera frame rate on the throughput. iPhone 6s supports to capture images with a wide range of frame rate from 3 to 240 fps, but the frame rate increases as the resolution decreases. We measure the achievable throughput by choosing a frame that achieves the highest frame throughput and multiplying it by the frame rate. The results are illustrated in Fig. 11c. The throughput reaches the maximum when the frames are captured in 3024p/30 format.

In CMOS cameras, the frame duration is limited by the readout duration $t_r$ and the resolution. As illustrated in Fig. 2, frame duration $t_f = (Y - 1)t_r + GAP$. Thus, the corresponding maximum frame rate is $1/t_f$. Hereafter, for the realistic performance evaluation, the capture format is set to 3024p/30.

### 5.3 Realistic Performance

In this section, we conduct our evaluations based on the following metrics:

- *Throughput*: the average amount of data successfully decoded per second in the received frames.
- *Bit error rate (BER)*: the percentage of wrongly decoded data in the total amount of data.
- *Overall error rate*: the percentage of overall reception error, including the frames failed to extract, the packets failed to demodulate, and the wrongly decoded bits. This can be expressed as $p_e = 1 - p_f \times p_p \times p_b$, where $p_f, p_p$ and $p_b$ are the percentage of successfully decoded frames, packets and bits, respectively.

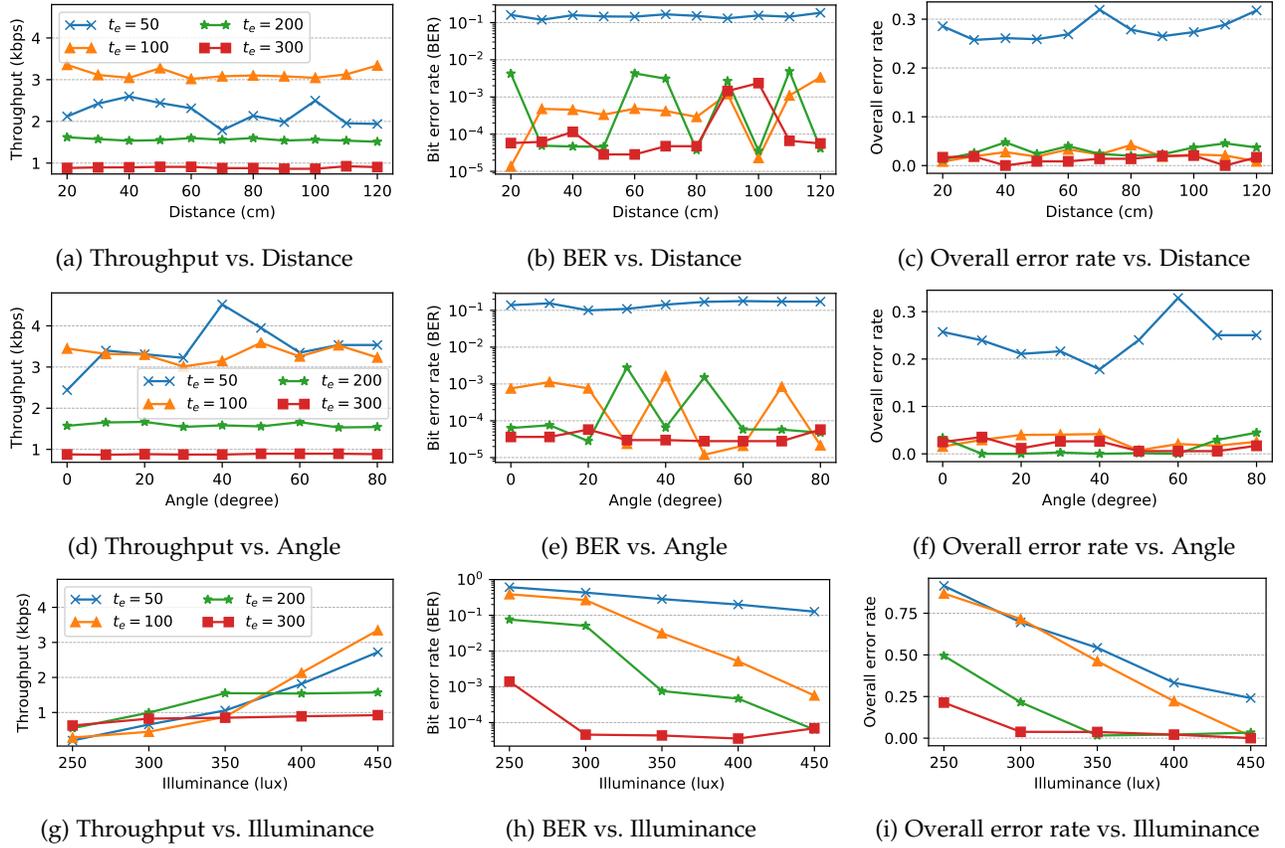

Fig. 13: The impacts of distance, angle, illuminance and exposure duration on throughput, BER and overall error rate, respectively. The reflector is 2.1 m away from the center LED. The illuminance is measured 0.5 m in front of the LED.

We examine the realistic performance under various environments with different exposure durations. To maximize the throughput potential of HYCACO, we choose three transmitters and four symbols per packet according to the evaluation results of achievable throughput in Section 5.2. We use iPhone 6s as the receiver. The resolution of the camera is set to $3024 \times 4032$, and the frame rate is 30 fps. We put a lumen meter 0.5 m in front of the center LED to measure the light level of the LED. When the LEDs are off, the meter reading is 50 lux.

#### 5.3.1 Effect of Transmitter Spacing

Since the channel gain differences among the LED-reflector links are crucial to the quality of recovered signals, we study how the adjacent distance between the transmitters affects the performance of HYCACO. The received light strength depends on its irradiance, incidence, and attenuation which follows a square law. We place the LEDs in line and the surfaces of LEDs parallel to the reflector, and the distance between the reflector and the LED surface is 2.1 m. We vary the adjacent distance between transmitters from 15 cm to 115 cm, while the receiver is fixed at a distance of 50 cm from the reflector. The time unit and the exposure duration is 100 $\mu s$. The measured illuminance is 450 lux. The experiment results are illustrated in Fig. 12. We can see that as the transmitter spacing increases, the throughput decrease, the BER and the overall error rate increase. The more precisely the illuminance proportionate to the number of ON-state LEDs, the higher quality of the reception will be. We hereafter fix the adjacent distance to 30 cm.

#### 5.3.2 Impact of Reflector-Receiver Distance

We evaluate the performance of HYCACO under a varying distance between the reflector and the receiver and report the results in Fig. 13a, 13b, 13c. The measured illuminance is 450 lux. The camera is parallel to the reflector. The results show that the throughput and reception error are independent of the distance. This is very different from the LoS LCC approaches, which are highly affected by the distance because the distance affects the RoI in the captured image. The limitation of the capturing distance is 120 cm because the reflector will not fill the field-of-view (FoV) of the camera if the distance further increases.

Images captured with low exposure durations will have low luminance and introduce more noise. Thus the signal layers have low SNRs. The shadow of the phone or the user on the reflector will further decrease the SNR which causes the fluctuation of the evaluation results. Therefore, the shorter exposure duration, the larger fluctuation.

#### 5.3.3 Impact of Viewing Angle

We then report in Fig. 13d, 13e, 13f the evaluation results under a varying viewing angle. Viewing angle is the angle between the camera and the reflector. Since the distance does affect the performance, we adjust the distance while varying the angle to make sure the reflector fill the FoV.

TABLE 2: Measured Illuminance, Voltage, and Current

| Illuminance (lux) | voltage (V) | Current (A) |
|---|---|---|
| 450 | 40 | 0.31 |
| 400 | 37.5 | 0.25 |
| 350 | 35.5 | 0.15 |
| 300 | 34 | 0.08 |
| 250 | 33 | 0.05 |

TABLE 3: Latency of Each Processing Step

| Processing Step | Latency (ms) |
|---|---|
| Sum pixels in each row and compute signal layer | 40.61 |
| Filter out DC component | 0.24 |
| Extract packets in a frame, Sampling, and normalize | 30.8 |
| Demodulate and decode a packet | 1.12 |

The measured illuminance is 450 lux. Similar to what we have observed with varying distances, the throughput and reception error are independent of the angle, too. Actually, the channel gain for each pixel sensor is different. That is why the frame has a brightest center and a gradual reduction of brightness on both sides (shown in Fig. 8), which makes it cannot directly set thresholds for detecting symbols. We address this challenge by capturing an extra long exposure image.

In summary, the reflector-receiver distance and the viewing angle have nearly no impact on the performance when the signal layer has sufficient SNR. But when the SNR is low, the shadow of the phone or the user, which is inevitable, may cause the fluctuation of the throughput. Furthermore, the viewing angle affects the position of the brightest center in a frame.

#### 5.3.4 Impact of Illuminance

Since the signal layer contains substantial noise for the high ISO, the performance is sensitive to the illuminance of the reflector. We evaluate the performance under a varying illuminance measured 0.5 m in front of the center LED. The receiver is fixed at a distance of 0.5 m from and parallel to the reflector. We vary the light output of the LEDs by adjusting the input currents, as shown in Table 2. The experiment results are reported in Fig. 13g, 13h, 13i. We can see that the plotted results of exposure durations below 100 $\mu s$ (included) have linear relationships as the illuminance increases; the plotted results of exposure durations above 100 $\mu s$ have linear relationships with the illuminance when the illuminance is below a threshold.

Long exposure duration helps the camera improve the quality of the captured image when the environment luminance is low, e.g., improve the image luminance, reduce the image noise. Therefore, if the reflector does not have a sufficient brightness, we can enhance the SNR by extending the exposure duration. However, the SNR is enhanced at a cost of reducing theoretical throughput.

#### 5.3.5 Impact of Exposure Duration

In the design of HYCACO, the exposure duration is the time unit of the duration of a symbol, which limits the theoretical throughput. We vary the exposure duration from 50 $\mu s$ to 300 $\mu s$, and the reported results show that the throughput increases as the exposure duration decreases. However, when the exposure duration and/or the illuminance is smaller than a threshold, the reception error increases which causes the throughput decreases. When the illuminance is higher than 450 lux and the exposure duration is longer than 100 $\mu s$, the frames are decoded with a BER lower than 1% and an overall error rate lower than 5%.

The supported range of exposure duration depends on the hardware of the experiment devices. Take the case of iPhone 6s, the duration of the signal recorded in an image is $(3024 - 1) \times 6.45 = 19489.35 \mu s$, which is the limitation of the packet duration. If the transmitter number is three and a packet has four symbols, to make sure a frame at least contains one complete packet, the maximum exposure duration is $19489.35 / [(6 \times 4 + 4) * 2] = 348.02 \mu s$. Although iPhone 6s allows us to set the exposure duration as short as 13 $\mu s$, it would result in a very low SNR. In the actual implementation of HYCACO, four main causes reduce the SNR:

- The latency of the operational amplifier in the switch circuit and the *ON-OFF* latency of the LED result in the slope of signal edge.
- For the smartphone cameras employ electronic shutters rather than mechanical shutters, the pixel sensor still accumulates the photons during the readout phase, which interferes the information for imaging.
- The brightness of a pixel is proportional to the amount of its received photons. Short exposure duration will result in low brightness.
- Summing up all the pixels in each row can reduce the image noise, but cannot eliminate the image noise.

### 5.4 Power Consumption and Latency

We finally measure the power consumption and the latency of processing $4032 \times 3024$ frames in our iOS receiver app. LCC is high power consumption, for CMOS sensors require a lot of power. We measure the energy impact with Xcode Instrument[2]. When the HYCACO app is capturing, processing and decoding simultaneously, the energy usage is $18 \pm 1$ (ranging between 0 and 20, with 20 indicating that the device is using power at a very high rate, and 0 indicating that very little power is being used). With this level of energy usage, the phone could run out the 1715 mAh battery within one or two hours. Furthermore, after about ten minutes of continuously receiving, the phone gets really hot.

The measured time consuming for each processing step is listed in Table 3. The processing steps are illustrated in Fig. 8. If the exposure duration (time unit) is 100 $\mu s$, the packet duration is 2800 $\mu s$. Five or six packets can be extracted in one frame. We assume that the user does not significantly move the phone, and the long exposure only needs to be captured once at the beginning of the reception. Thus, the total processing time of each frame is about 78 ms. The major cause of the latency is due to the high resolution, which contains twelve million pixels. If we lower the resolution by half, the processing can be done in realtime. Alternatively, we can reduce the latency by summing only part pixels in each row into a sample, but this will bring down the SNR.

---

2. The Energy Usage in Xcode Instrument indicates a level from 0 to 20, indicating how much energy the app is using at any given time. These numbers are subjective.

## 6 DISCUSSION

Cameras on the market usually have different frame rates, resolution, and readout durations. The throughputs of LCC systems are highly related to these features. Moreover, LCC is unsuitable for continuous receptions due to its high power consumption. Therefore, the first aim of an LCC system should be easy to use rather than the high throughput. The application scenarios of LoS LCC are restricted by the small RoI and the FoV of the camera. Light suffers less from multipath effects than WiFi signals, and the radiant intensities follow an attenuation pattern. NLoS LCC can be used for indoor positioning and/or orientation techniques by estimating the channel gain of the transmitter-reflector links. We plan it for future work.

Our prototype works fine under a condition that the illuminance perceived by the camera is proportional to the number of ON-state LEDs. Employing more LED luminaires for transmission may violate the assumption. The relative position of the camera and the reflector has nearly no impact on the performance, because the channel gain of the reflector-receiver link is the same for all transmitters. Whereby the perceived illuminance differences are dominated by the transmitter-reflector links. When HYCACO can work with channel compensations, the application scenarios will be various.

## 7 RELATED WORK

**NLoS LCC.** Most of the existing NLoS LCC approaches only utilize one LED for transmission. Danakis et al. [13] first propose that the CMOS camera can be used as a receiver in order to capture the continuous changes of the status (ON-OFF) of the light. MILC [14] improves the throughput with multi-level illuminations. Martian [3] encodes bits by varying the duty cycle of a pulse waveform. Rajagopal et al. [15] propose a hybrid VLC system, which simultaneously transmits low-speed data to cameras and high-speed data to photodiode receivers. ReflexCode [4] adopts reflected light emitted from multiple LEDs as its communication media, which looks very similar to our work, but it is actually another way to implement multi-level illuminations. Using a single LED to provide multi-level illuminations need an additional hardware like a DAC, which will increase the cost. Varying the luminous intensity or the duty cycle may break the overall brightness energy balance. The LEDs in HYCACO flicker with a constant duty cycle (50%) and thus are naturally flicker-free to human eyes.

**LoS LCC.** If an LED has sufficient brightness, the LoS link has a sufficient SNR and thus is more resilient to the ambient noise. RollingLight [2] employs frequency shift keying scheme and delivers a throughput of 11.32 Bps. However, this throughput is achieved when the camera is very close to the LED. CamCom [16] uses undersampled frequency shift OOK to encode bits, and it achieves a throughput of 400 bps using 100 LEDs. Luo et al. [17] propose undersampled phase shift OOK, and their system reaches 150 bps with a dual LED lamp. ColorBars [18] utilizes Color Shift Keying (CSK) to modulate data using different colors transmitted by the LED. The major bottleneck of throughput is the small RoI. Besides, LoS LCC provides a natural way to enable visual association, which creates an opportunity for indoor positioning. Luxapose [19] explores the indoor positioning problem by detecting the presence of the luminaires in the captured image. LiTell [20] proposes a robust localization scheme that employs unmodified fluorescent lights (FLs) as location landmarks. However, these positioning schemes need LoS links and enough spatial resolutions to separate transmissions from different transmitters.

**Other Recent VLC Works.** Several recent works investigate other specific types of VLC. Screen-to-camera communications [8], [21], [22] employ the screens as the transmitters which have larger resolution and more changing states. Disco [7] uses a modified display as the transmitter to send sine wave signals. It recovers the signals with a simultaneous dual exposure (SDE) sensor. DarkLight [23] and DarkVLC [24] allows light-based communication to be sustained even when the LED lights appear dark or OFF. Kaleido [9] utilizes the rolling shutter effect to prevent unauthorized users from taping a video played on a screen. LCC is a just a special type of VLC. In the transmitter, the rolling shutter deforms the signal frequency spatial detail by convolution. Thereby, with the proposed signal recovery approach, lots of modulation schemes employed in other kinds of VLC systems can also be employed in LCC.

## 8 CONCLUSION

In this paper, we design and implement HYCACO, which enables multiple access for NLoS LCC. We demonstrated the efficacy of our design using a hardware prototype, which achieves a throughput of 4.5 kbps. HYCACO works fine when the measured illuminance is 450 lux which is the recommended illuminance for an indoor environment. Unlike the width-driven demodulation [2], [3], which is complex in computation, we just sample hundreds of judging points from the received signal for the demodulation. Therefore, HYCACO is high computational efficiency and can provide realtime responses with hand-held devices. The major concern of NLoS links is the signal power attenuation. We extract the signal layer from the short exposure image by dividing it from a long exposure image. Thus, not only the image background is eliminated, but also the SNR is enhanced. If we use the existing LED infrastructures for transmissions, the channel gains of the transmitter-reflector links will be different. We can add a compensation algorithm to make HYCACO functional or use the channel properties for indoor positioning. We plan to address these challenges in future work.

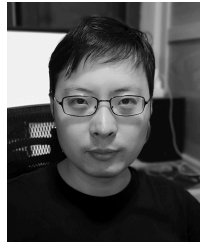

**Fan Yang** received the BS degree in computer science from Qingdao University, Qingdao, China, in 2009. He received the MS degree in software engineering from Xidian University, Xi'an, China, in 2014. He is currently working toward the PhD degree with the School of Computer Science at Northwestern Polytechnical University. His research interests include indoor positioning and visible light communications.

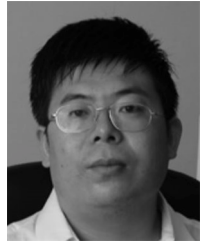

**Shining Li** received the BS and MS degrees in computer science from Northwestern Polytechnical University, Xi'an, China, in 1989 and 1992, respectively. He received the PhD degree in computer science from Xi'an Jiaotong University, Xi'an, China, in 2005. He is currently a professor at the School of Computer Science, Northwestern Polytechnical University. His research interests include mobile computing and wireless sensor networks. He is a member of the IEEE.

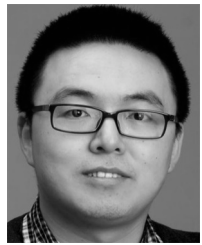

**Zhe Yang** received his B.S. degree in information engineering in 2005 and the M.S. degree in control theory and engineering in 2008, both from Xi'an Jiaotong University, Xi'an, China. He received the Ph.D. degree in electrical and computer engineering from the University of Victoria, Victoria, British Columbia, Canada, in 2013. He then joined the Department of Computer Science at Northwestern Polytechnical University, Xi'an, China, with the exceptional promotion to associate professor. His research areas include protocol design, optimization, and resource management of wireless communication networks.

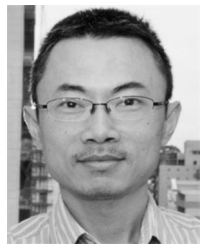

**Tao Gu** is currently an associate professor in computer science with RMIT University, Australia. His current research interests include mobile computing, ubiquitous/pervasive computing, wireless sensor networks, distributed network systems, sensor data analytics, cyber physical system, Internet of Things, and online social networks. He is a senior member of the IEEE and a member of the ACM.

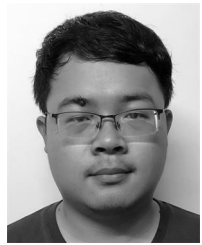

**Cheng Qian** received the BE degree in Internet of things Engineering from Northwestern Polytechnical University, China in 2017. Since October 2017, he has been working toward the MS degree in Computer Science and Technology from Northwestern Polytechnical University.